\documentclass[10pt,draft]{dis03}
\usepackage{epsf,amsmath}

\def\gsim{\,\lower.25ex\hbox{$\scriptstyle\sim$}\kern-1.30ex%
\raise 0.55ex\hbox{$\scriptstyle >$}\,}
\def\lsim{\,\lower.25ex\hbox{$\scriptstyle\sim$}\kern-1.30ex%
\raise 0.55ex\hbox{$\scriptstyle <$}\,}

\newcommand{\ra}{\rightarrow}
\newcommand{\ccb}{c\bar{c}}
\newcommand{\bbb}{b\bar{b}}

\newcommand{\epem}{e^+e^-}

\newcommand{\agt}{\,\rlap{\lower 3.5 pt \hbox{$\mathchar \sim$}} \raise 1pt
 \hbox {$>$}\,}
\newcommand{\alt}{\,\rlap{\lower 3.5 pt \hbox{$\mathchar \sim$}} \raise 1pt
 \hbox {$<$}\,}

\begin{document}

\title{Summary of the Heavy Flavour Session
 \thanks{To appear in the proceedings of the 
     XI International Workshop on Deep Inelastic Scattering (DIS'03),
St.\ Petersburg, April 2003}
}

\author{B.A. KNIEHL \\
II. Institut f\"ur Theoretische Physik, Universit\"at Hamburg,\\
Luruper Chaussee 149, 22761 Hamburg, Germany\\ \\ 
F. SEFKOW \\
DESY, \\
Notkestr.\ 85, 22607 Hamburg, Germany}

\maketitle

\begin{abstract}
\noindent The contributions to the 
parallel session on heavy flavour production are summarized. 
In more than 30 presentations new theoretical developments and 
new experimental results from the Tevatron, HERA, LEP, the B factories
and $\nu N$ scattering were reported.   
\end{abstract}

\section{Introduction} 

Although traditionally centered around deep-inelastic scattering (DIS) and
structure functions, this conference series has become one of the
prominent places to discuss strong interaction physics in a wider sense.
In the heavy flavour case in particular, the focus is on understanding
in a common QCD framework the 
charm, beauty and top cross sections in various production environments.

The experimental results reported in this session were obtained in $pp$, $pN$,
$\nu N$, $ep$, $\gamma p$, $\gamma\gamma$ and $\epem$ collisions. 
Only a personal and biased selection from them can be mentioned 
in this summary.
Since the status of heavy flavour physics at HERA has been well covered in the 
introductory talk~\cite{naroska}, the emphasis is here on new results and
on other machines. 

The theoretical results presented in this session capture various recent
developments in the area of inclusive charmonium, open-charm, and open-bottom
production, as well as the heavy-flavour contributions to the structure
functions \cite{tun,oln}.

\section{Charmonium (Theory)}
\label{sec:oniumtheo}

Since the discovery of the $J/\psi$ meson in 1974, heavy quarkonium has
provided a useful laboratory for quantitative tests of QCD and, in particular,
of the interplay of perturbative and nonperturbative phenomena.
The factorization formalism of nonrelativistic QCD (NRQCD) \cite{bbl} provides
a rigorous theoretical framework for the description of heavy-quarkonium
production and decay.
This formalism implies a separation of short-distance coefficients, which can
be calculated perturbatively as expansions in the strong-coupling constant
$\alpha_s$, from long-distance matrix elements (MEs), which must be extracted
from experiment.
The relative importance of the latter can be estimated by means of velocity
scaling rules; i.e., the MEs are predicted to scale with a definite power of
the heavy-quark ($Q$) velocity $v$ in the limit $v\ll1$.
In this way, the theoretical predictions are organized as double expansions in
$\alpha_s$ and $v$.
A crucial feature of this formalism is that it takes into account the complete
structure of the $Q\overline{Q}$ Fock space, which is spanned by the states
$n={}^{2S+1}L_J^{(a)}$ with definite spin $S$, orbital angular momentum
$L$, total angular momentum $J$, and color multiplicity $a=1,8$.
In particular, this formalism predicts the existence of color-octet (CO)
processes in nature.
This means that $Q\overline{Q}$ pairs are produced at short distances in CO
states and subsequently evolve into physical, color-singlet (CS) quarkonia by
the nonperturbative emission of soft gluons.
In the limit $v\to0$, the traditional CS model (CSM) \cite{ber} is recovered.
The greatest triumph of this formalism was that it was able to correctly
describe \cite{bra} the cross section of $p\overline{p}\to J/\psi+X$ at the
Tevatron \cite{abe}, which had turned out to be more than one order of
magnitude in excess of the theoretical prediction based on the CSM.

Apart from this phenomenological drawback, the CSM also suffers from severe
conceptual problems indicating that it is incomplete.
These include the presence of logarithmic infrared divergences in the
${\cal O}(\alpha_s)$ corrections to $P$-wave decays to light hadrons and in
the relativistic corrections to $S$-wave annihilation \cite{bar}, and the lack
of a general argument for its validity in higher orders of perturbation
theory.
In the case of $\gamma p\to J/\psi+X$ in photoproduction, where
next-to-leading (NLO) corrections in the CSM happen to be well defined
\cite{zun}, the NLO prediction encompasses the rather precise
HERA data \cite{gammap} within the theoretical uncertainties.

In order to convincingly establish the phenomenological significance of the CO
processes, it is indispensable to identify them in other kinds of high-energy
experiments as well, using the very MEs extracted from the Tevatron data
\cite{abe}.
Attempts to do so range from very successful to inconclusive for the time
being.
Successful examples include the comparison of theoretical predictions of
$\gamma\gamma\to J/\psi+X$ \cite{gg} and $pp\to J/\psi+X$ \cite{pol} with data
taken by DELPHI \cite{delphi} at LEP2
and by PHENIX \cite{adl} at RHIC, respectively.
Comparison of HERA \cite{h1} data of $ep\to J/\psi+X$ in DIS with NRQCD
predictions \cite{ep} are generally favourable with respect to both 
normalization and shape, except for the distribution in inelasticity $z$,
where a shape problem familiar from photoproduction \cite{cac} is recovered.
As for the $J/\psi$ polarization in $p\overline{p}\to J/\psi+X$, measurements
at the Tevatron \cite{aff} tend to undershoot the theoretical predictions for
direct \cite{ben} and prompt \cite{bkl} production 
in the bins of highest transverse momentum ($p_T$).

While the $k_T$-factorization \cite{sri,spb} and hard-comover-scattering
\cite{hoy} approaches manage to bring the CSM prediction of
$p\overline{p}\to J/\psi+X$ much closer to the Tevatron data \cite{abe}, they
do not cure the conceptual defects of the CSM; NRQCD is still needed for that.
However, they reduce the relative importance of the CO contributions at
leading order (LO).
Since NLO corrections allow for an incoming parton to gain a $k_T$ kick by
emitting another parton before entering the hard scattering, one expects
$k_T$-factorization effects to become much less important at NLO.
While the evaluation of NLO corrections is well understood in NRQCD, the
extension of the $k_T$-factorization formalism beyond LO is an open problem.
The list of open problems in connection with the $k_T$-factorization formalism
also includes the determination of the proper evolution equation (BFKL, CCFM,
DGLAP, or mixtures thereof), the establishment of a rigorous operator 
definition of the unintegrated parton density functions, the restoration of
gauge independence for off-shell external partons, the avoidance of ad-hoc
regularization to keep propagators away from poles, which are sometimes used 
in the literature \cite{spb}, and the recovery of the collinear-parton-model
results for asymptotically large values of $p_T$ in practical applications
\cite{spb}.

The color evaporation model \cite{cem} provides an operative procedure to
estimate the production cross sections of heavy quarkonia on the basis of the
ones of open heavy-quark pairs.
It is intuitive and useful for qualitative studies, and it also significantly
improves the description of the Tevatron data as compared to the CSM
\cite{sch}.
However, it does not account for the process-specific weights of the CS and CO
contributions, but rather assumes a fixed ratio of $1:7$.

\section{New Charmonium measurements}

The analysis of inelastic $J/\psi$ production with HERA I data is essentially
completed. At this meeting, ZEUS presented preliminary cross section results 
for charmonium production in DIS~\cite{katkov} 
which confirm last year's H1 findings, namely that NRQCD with sizeable 
CO contributions reproduces 
the data reasonably well, in particular at higher $p_T$. However, 
the dependence on the inelasticity variable $z$ remains at variance with 
the calculations also in the new ZEUS data. 

New experimental testing grounds may provide further clues to clarify 
the validity of the NRQCD factorization approach. 
In addition to the $\gamma\gamma$ collision results~\cite{delphi} 
(Sec.~\ref{sec:oniumtheo}), 
data on charmonium production measured by HERA-B, at the B factories in the    
$\epem$ continuum, or at the Tevatron in the low $p_T$ regime are entering 
the stage. 

The HERA-B collaboration has published a first measurement of $R_{\chi_c}$, 
the fraction of $J/\psi$ mesons produced via intermediate $\chi_c$ states, 
which lies considerably below CSM expectations, but matches the NRQCD 
predictions well~\cite{kruecker}. 
The total statistics of about 3000000 $J/\psi$ and 20000 $\chi_c$ 
mesons produced on different nuclear targets allows them, 
for example, to
measure the cross section dependence on isotope number $A$.
In NRQCD this is not the same for $J/\psi$ and $\chi_c$, since 
CS and CO pre-resonant states contribute with different strength and 
interact differently with nuclear matter~\cite{rvogtadep}. 
A measurement of the $A$ dependence of the $\chi_c$ cross section also
forms an important ingredient for the interpretation of 
heavy ion collision data.

\begin{figure}[htb]
  \begin{picture}(10,7)(0,0)
\includegraphics{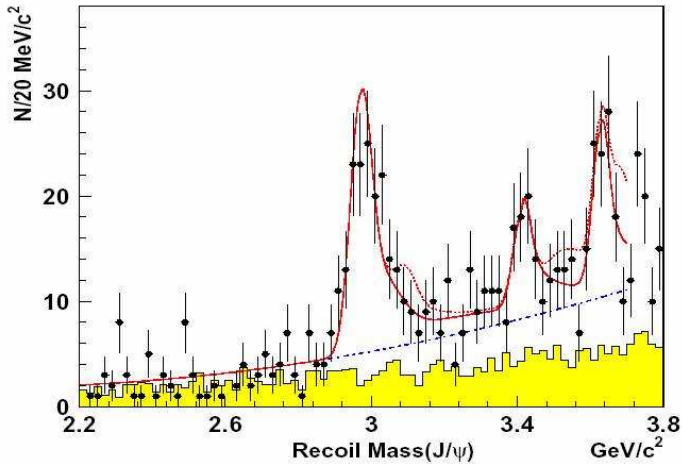}
\end{picture}
\caption{\label{fig:psipsi}
Invariant mass of the system recoiling against $J/\psi$ mesons produced in 
$\epem$ annihilation below the Upsilon resonance, measured by Belle.}
\end{figure}

A rather new testing field is charmonium production in the $\epem$ 
annihilation continuum below the $\Upsilon (4S)$ resonance~\cite{pakhlov}.
The cross section was found to be  more than an order of magnitude 
higher than expected, but the momentum spectrum does not exhibit an 
enhancement at large momenta, which would signal strong CO contributions.  
According to the preliminary Belle results, $(82\pm 20)\%$ of 
the $J/\psi$ mesons are accompanied by another pair of 
charm quarks. 
The $J/\psi$ recoil mass spectrum, measured now in a sample corresponding 
to 101 pb$^{-1}$ and shown in Fig.~\ref{fig:psipsi}, exhibits several clear 
$\ccb$ resonances. 
\unitlength1cm
From these data, rates for many exclusive (double) 
charmonium modes were be 
measured, or earlier measurements confirmed with better precision; 
in addition all exclusive open charm ground states have been observed 
recoiling against $J/\psi$ mesons. 

New possibilities also open up in familiar environments. 
The Tevatron experiments are vigorously entering back on stage 
with new data from run II. By the time of the DIS conference, 
CDF, for example, wrote 150~pb$^{-1}$ to tape and presented analyses making 
use of about 70~pb$^{-1}$, to be compared to the run I data set corresponding 
to 110~pb$^{-1}$.
With their upgraded detector, trigger and data acquisition systems 
kinematic ranges and channels can be probed which have been inaccessible
so far in highest energy $\bar{p}p$ collisions~\cite{meyer}.
One example are lowered dimuon trigger thresholds which allow to trigger 
on $J/\Psi$ mesons decaying at rest and thus to measure their production 
cross section over the entire momentum range down to 
$p_T=0$. 
The existing measurements of run I start at $p_T>5$~GeV and 
cover less than 10\% of the total cross section.
This allows one to test the NRQCD description over a wider range; it 
provides increased sensitivity to those CO MEs
($^1S_0$ and $^3P_J$) which also govern $J/\psi$ production at HERA 
and should enable more stringent tests of their universality. 
However, a complete picture and a meaningful statement on the validity of 
the NRQCD factorization approach will only emerge once the NLO 
corrections, which are shown to be important in the HERA photoproduction case,
become available for hadroproduction as well. This requires a huge effort
on the theoretical side, which should bear fruit in the not so far future.
 
\section{Open heavy flavours (Theory)}

The QCD-improved parton model implemented in the $\overline{\rm MS}$
renormalization and factorization scheme and endowed with nonperturbative
fragmentation functions (FFs), which proved itself so convincingly for
light-meson inclusive production \cite{kkp}, also provides an ideal
theoretical framework for a coherent global analysis of $D$- \cite{bkkc} and
$B$-meson data \cite{bkkb}, provided that $\mu\gg m_Q$, where $\mu$ is the
energy scale characteristic for the respective production process and $Q=c,b$.
Then, at LO (NLO), the dominant logarithmic terms, of the form
$\alpha_s^{n,n+1}\ln^n\left(\mu^2/m_Q^2\right)$ with $n=1,2,\ldots$, are
properly resummed to all orders by the DGLAP evolution, while power terms of
the form $\left(m_Q^2/\mu^2\right)^n$ are negligibly small and can be safely
neglected.
In this {\it massless-quark scheme} or {\it zero-mass variable-flavour-number
scheme} (ZMVFNS), which is sometimes quite wrongly referred to as {\it NLL
approximation},\footnote{%
The nonlogarithmic corrections of relative order $\alpha_s$ are fully
included, except for terms that are suppressed by powers of
$\left(m_Q^2/\mu^2\right)^n$.}
the $Q$ quark is treated as massless and appears as an active parton in the
incoming hadron or photon, having a nonperturbative parton density function.
The criterion $\mu\gg m_Q$ is certainly satisfied for $e^+e^-$ annihilation on
the $Z$-boson resonance, and for the photo-, lepto-, and hadroproduction of
$D$ and $B$ mesons with $p_T\gg m_Q$.
Furthermore, the universality of the FFs is guaranteed by the factorization
theorem \cite{col}, which entitles us to transfer information on how $c$ and 
$b$ quarks hadronize to $D$ and $B$ mesons, respectively, in a well-defined
quantitative way from $e^+e^-$ annihilation, where the measurements are
usually most precise, to other kinds of experiments, such as photo-, lepto-, 
and hadroproduction.
In Refs.~\cite{bkkc,bkkb}, the distributions in the scaled $D$- and $B$-meson
energy $x=2E/\sqrt s$ measured at LEP1 were fitted at LO and NLO using, among
others, the ansatz by Peterson et al.\ \cite{pet} for the $c\to D$ and
$b\to B$ FF at the starting scale $\mu_0=2m_Q$.
In the $D^{\star\pm}$ ($B^+/B^0$) case, the $\varepsilon$ parameter was found
to be $\varepsilon_c=0.0851$ and $0.116$ \cite{bkkc} ($\varepsilon_b=0.0126$
and $0.0198$ \cite{bkkb}) at LO and NLO, respectively.
We emphasize that the value of $\varepsilon$ carries no meaning by itself, but
it depends on the underlying theory for the description of the fragmentation
process, in particular, on the choice of the starting scale $\mu_0$, on
whether the analysis is performed in LO or NLO, and on how the final-state
collinear singularities are factorized in NLO.

In the traditional {\it massive-quark scheme} or {\it fixed-flavour-number
scheme} (FFNS), the $Q$ quark is treated in the on-mass-shell renormalization
scheme, as if it were a massive lepton in triplicate, and it only appears in
the final state, but not as an active parton inside the incoming hadron or
photon.
There are no collinear singularities associated with the outgoing $Q$-quark
lines that need to be subtracted and absorbed into FFs.
In fact, in this scheme, there is no conceptual necessity for FFs at all, but
they are nevertheless introduced in an ad-hoc fashion in an attempt to match
the $D$- and $B$-meson data.
However, in the absence of a subtraction procedure, there is also no
factorization theorem in operation to guarantee the universality of the FFs
\cite{col}.
By the same token, such FFs are not subject to DGLAP evolution.
Thus, there is no theoretical justification to expect, e.g., that a single
value of the Peterson $\varepsilon$ parameter should be appropriate for
different types of experiment or at different energy scales in the same type
of experiment.
In other words, the feasibility of global data analyses is questionable in
this scheme.
Moreover, this scheme breaks down for $p_T\gg m_Q$ because of would-be
collinear singularities of the form $\alpha_s\ln\left(p_T^2/m_Q^2\right)$,
which are not resummed.
However, it allows one to calculate a total cross section, which is infeasible
in the ZMVFNS.

The attempt to split the $D$- and $B$-meson FFs into so-called perturbative
FFs (PFFs) and nonperturbative remainders is interesting in its own right.
However, detailed analysis for $D^{*\pm}$-meson FFs \cite{pff} revealed that
such a procedure leads to deficient results in practical applications.
On the one hand, at NLO, the cross section $d\sigma/dx$ of $e^+e^-$
annihilation becomes negative in the upper $x$ range, at $x\agt0.9$, where the
data is very precise, so that a low-quality fit is obtained unless this $x$
range is excluded by hand \cite{pff,gre}.
On the other hand, the LO and NLO predictions for other types of processes,
such as photoproduction in $ep$ scattering, significantly differ \cite{pff},
which implies that the perturbative stability might be insufficient.

The idea \cite{bkkb} of performing a coherent analysis of LEP1 and Tevatron
data of inclusive $B$-meson production was recently revived using an
unconventional scheme named {\it fixed-order next-to-leading-logarithm}
(FONLL) scheme, in which the traditional result in the FFNS and a suitably
subtracted result in a ZMVFNS with PFFs are linearly combined \cite{cn}.
However, some degree of arbitrariness is inherent to  this procedure, 
as may be understood by noticing that
the ZMVFNS term is weighted with an ad-hoc coefficient function of the form
$p_T^2/\left(p_T^2+25m_b^2\right)$ so as to effectuate its suppression in the
low-$p_T$ range and that this term is evaluated at
$p_T^\prime=\sqrt{p_T^2+m_Q^2}$ while the FFNS term is evaluated at $p_T$.
Since the FONLL scheme interpolates between the FFNS and the ZMVFNS with PFFs,
it also inherits some weaknesses of both schemes detailed above.
In particular, the negativity of the NLO cross section of $e^+e^-\to D/B+X$ in
the upper $x$ range forces one to exclude the data points located there from
the fit.
In Ref.~\cite{cn}, this is achieved by resorting to what is called there the
{\it moments method}, i.e., the large-$x$ region is manually faded out by
selecting one particular low moment of the FF, namely the one corresponding to
the average $x$ value, thereby leaving the residual information encoded in the
data unused.

A rigorous theoretical framework that retains the full finite-$m_Q$ effects
while preserving the indispensible virtues of the factorization theorem,
namely the universality and the DGLAP scaling violation of the FFs, is
provided by the {\it general-mass variable-flavour-number scheme} (GMVFNS)
\cite{col,acot}.
In a nut shell, this procedure consists in explicitly performing the $m_Q\to0$
limit of the FFNS result, comparing the outcome, term by term, with the
ZMVFNS result in the $\overline{\rm MS}$ scheme, and subtracting the
difference terms from the FFNS result.
Owing to the factorization theorem \cite{col}, the hard-scattering cross
sections thus obtained can then be convoluted with nonperturbative $D$- and
$B$-meson FFs extracted from LEP1 data using the $\overline{\rm MS}$ scheme
\cite{bkkc,bkkb}.
This is consistent because the finite-$m_Q$ terms omitted in
Refs.~\cite{bkkc,bkkb} are negligibly small, of order $m_Q^2/m_Z^2$.
In fact, a reanalysis of the LEP1 data in the GMVFNS should yield FFs that
agree with the FFs of Refs.~\cite{bkkc,bkkb} within the errors.
The GMVFNS was recently implemented for direct \cite{ks1} and single-resolved
\cite{ks2} $\gamma\gamma$ collisions. 
In the case of $\gamma\gamma\to D^{\star\pm}+X$ at LEP2, the inclusion of
finite-$m_c$ effects was found to reduce the cross section by approximately
20\% (10\%) at $p_T=2m_c$ ($3m_c$) \cite{ks1}, i.e., their magnitude is
roughly $m_c^2/p_T^2$, as na\"\i vely expected.
By analogy, one expects the finite-$m_b$ terms neglected in Ref.~\cite{bkkb}
to have a moderate size, of order 20\% (10\%) at $p_T=10$~GeV (15~GeV).
This is considerably smaller than the scale uncertainty and appears
insignificant compared to the excess of the CDF data \cite{cdf} over the
traditional NLO analysis in the FFNS \cite{nas}.

The GMVFNS scheme can be further improved by implementing the proper kinematic
constraints on the threshold behaviour.
In the case of the structure function $F_2^c$, this was done in
Ref.~\cite{tun} by appropriately introducing the variable
$\chi=x\left(1+4m_c^2/Q^2\right)$ (ACOT($\chi$) prescription).
This was shown to enforce the proper threshold behaviour, to simplify the
implementation, and to yield robust predictions.

\section{Heavy quark heavy hadron transition}

In a number of talks the current understanding of heavy quark fragmentation
(or hadronization) was reviewed. This enters into every comparison of 
production cross sections with theory, to be discussed in the subsequent 
section.
 
\begin{figure}[htb]
  \begin{picture}(10,7)(0,0)
\includegraphics{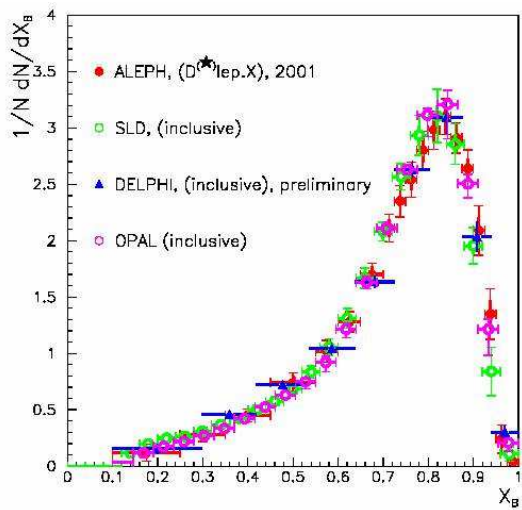}
\end{picture}
\caption{\label{fig:bfrag}
Fractional momentum distribution of the weakly decaying $b$ hadron, 
measured in $\epem$ annihilation at the Z resonance.}
\end{figure}

Above all, fragmentation is very well constrained by experiment. 
The most precise data are nowadays those on B meson production in 
$\epem$ annihilation at the Z resonance~\cite{kerzel,landsman}.
The compilation of LEP and SLD measurements in Fig.~\ref{fig:bfrag}
shows impressive agreement between the different analyses, 
which are based on exclusive ($D^*$ lepton), semileptonic or inclusive 
vertex methods to tag the $b$ hadron and infer its momentum from the 
partially reconstructed final state.    
\unitlength1cm
The figure displays the unfolded spectrum of the momentum fraction 
of the weakly decaying hadron, normalized to the beam energy. 
It reflects both the effects of the hard QCD radiation processes and 
of the soft transition from the quark to the bound hadronic state. 
The experiments have also studied how well the data can be described when 
different parameterizations are used to model the soft phase, and arrive
at the same ranking~\cite{harder}, namely that the Lund and the Bowler shapes
are preferred with respect to, e.g. the Peterson parameterization.  
Of course this conclusion is relevant only when a leading order (LO) 
parton shower approach as it is implemented in the JETSET Monte Carlo 
program, is used to model the perturbative phase.

At HERA, the charm fragmentation function has been probed~\cite{gladilin}.
Since the initial state does not constrain the primordial quark momentum, 
the beam energy has been replaced by the jet energy to normalize 
the charm hadron's momentum fraction $x$. 
Hard gluon radiation thus affects the observable $x$ spectrum less than 
in the $\epem$ case. The measured distributions obtained 
by ZEUS, ARGUS and OPAL, 
which are of course not expected to agree in detail, exhibit nevertheless 
the same gross features,
and fitting a parton shower Monte Carlo prediction 
to the ZEUS data yields a Peterson parameter 
consistent with $\epem$ findings. 

An enhancement at low $x$, attributed to gluon splitting, is observed only 
in the OPAL data, in the ARGUS and ZEUS case the energies are too low. 
Gluon splitting to $\ccb$ and $\bbb$ pairs has been explicitly constrained 
by studying special (3 and 4 jet) event topologies 
at LEP~\cite{landsman,giammanco}.
For charm, a new and quite precise ALEPH measurement  
and a new world average of $g_{cc}=(3.01\pm0.33)\%$ has been presented 
at this conference~\cite{giammanco}. 
This is somewhat above theoretical expectation, 
but consistent within errors.
The situation is similar for the ten times smaller 
$g_{bb}$ value~\cite{landsman}; 
altogether this shows that also higher order corrections 
are well under control.

\begin{figure}[htb]
  \begin{picture}(10,5)(0,0)
\includegraphics{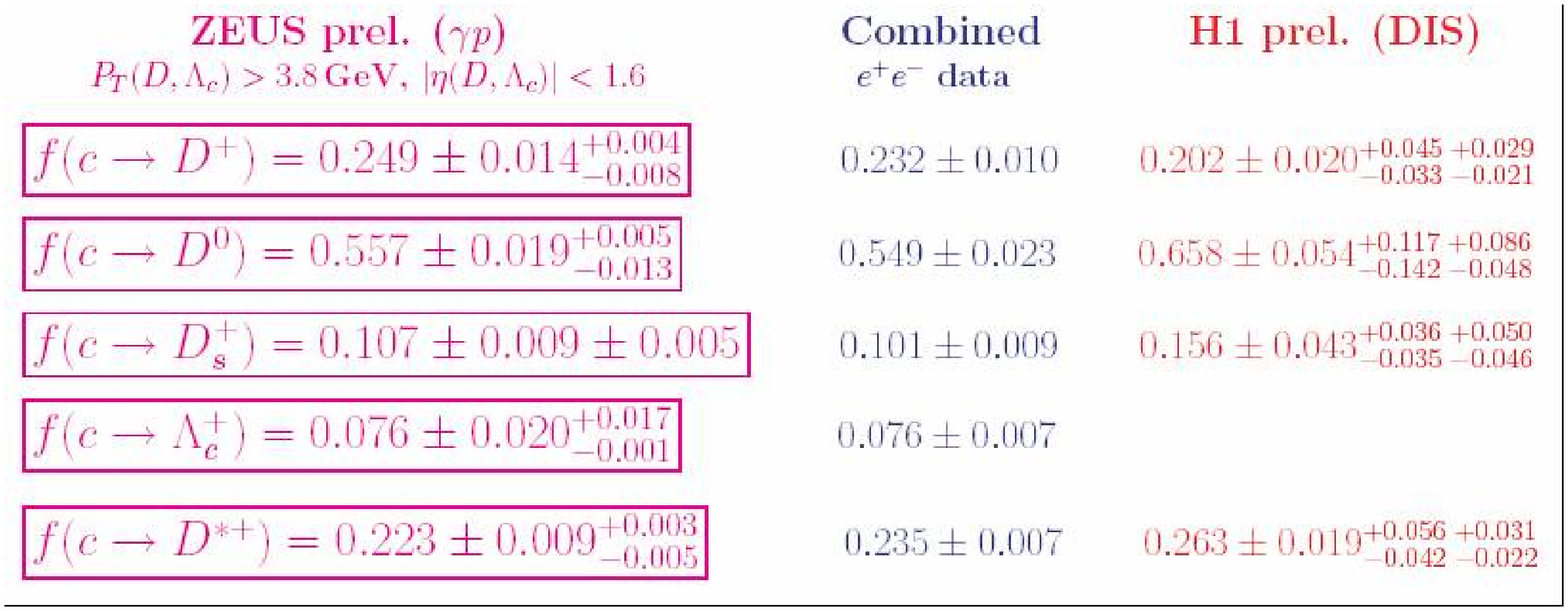}
\end{picture}
\caption{\label{tab:fragfrac}
Fragmentation fractions for charm hadrons.}
\end{figure}

Another way to test the universality of the non-perturbative 
quark-hadron transition is to compare the 
fragmentation fractions~\cite{gladilin}, 
i.e.\ the relative rates with which the different weakly decaying heavy hadron 
states are being produced.
ZEUS measured now, in addition to $D^*$, $D^0$ and $D_s^+$, 
also the production of $D^+$ and $\lambda_c$ 
(to represent the charmed baryons) 
and have thus determined the complete set, 
so that the rates can now be normalized to the sum 
and no recurrence to MC for the total charm cross section has to be made
anymore. 
Table~\ref{tab:fragfrac} compares the ZEUS results with H1 measurements, 
showing consistency of the HERA data among each other,
and with $\epem$ results, which confirms the universality of 
characteristics like the occurrence of strange mesons, baryons or 
vector mesons. 
\unitlength1cm
These results further validate the practice to use the clean and precise 
measurements from $\epem$ annihilation 
to model and interpret $ep$ data.
They give experimental support to
a rigorous scheme to relate the QCD theory of quarks 
to observable single hadrons, which was sketched above and 
which provides a powerful means to probe parton dynamics.

\section{Open $c$, $b$ and $t$ production} 

From an industrious community at HERA and other machines,
cross section data on open heavy flavour production keep coming in 
and providing new challenges to the common QCD framework.

\begin{figure}[htb]
  \begin{picture}(10,8)(0,0)
\includegraphics{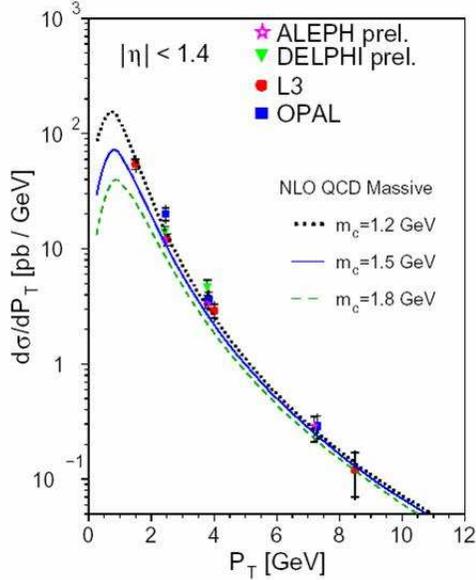}
\end{picture}
\caption{\label{fig:ggdst}
$D^*$ cross section in two photon interactions at LEP2,
compared with fixed order (massive) QCD calculations.}
\end{figure}

The two-photon analyses at LEP2 are being finalized.
$\bbb$ production, after the L3 and OPAL observations, 
is now also seen by DELPHI~\cite{achard}
to be in excess of the massive NLO QCD prediction by more than a factor of 2.
Direct and resolved photon processes contribute here with similar strength. 
The result appears to be in contrast to the two-photon data 
on charm production which are reproduced by theory, 
within errors. 
An inspection of the $p_T$ spectrum (Fig.~\ref{fig:ggdst})
reveals that data lie rather at the upper side 
of the error range of the theoretical prediction, 
in particular for small transverse momenta. 
\unitlength1cm
One should bear in mind that the $\bbb$ data are dominated by 
final states with $p_T\sim m_b$, populating a region where
the factorization ansatz may be at the limit of its validity. 

\begin{figure}[htb]
  \begin{picture}(10,8)(0,0)
\includegraphics{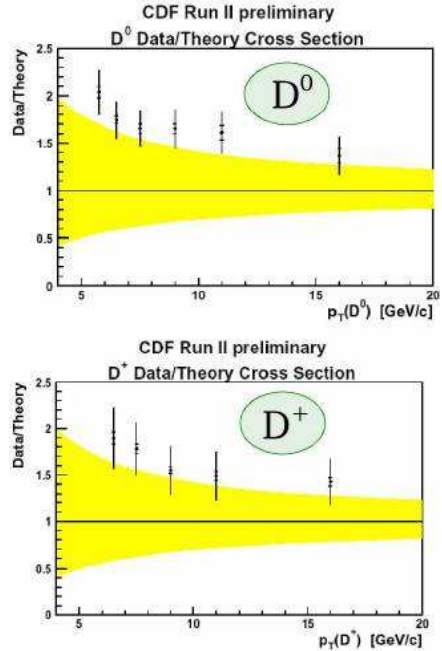}
\end{picture}
\caption{\label{fig:cdfd}
Charm meson production rate at the Tevatron, normalized to resummed 
QCD predictions in the FONLL scheme.}
\end{figure}

Thanks to their upgraded trigger capabilities, CDF was able to present
the first open charm cross section data from the Tevatron 
to this conference~\cite{meyer}.
The results, presently extracted from a rather small amount of luminosity,
are shown in Fig.~\ref{fig:cdfd} in comparison with resummed QCD calculations
in the FONLL scheme.. 
\unitlength1cm
The data tend to be above the central prediction, but come closer as 
$p_T$ increases and thus follow the behaviour of  
the theoretical scale uncertainty.

\begin{figure}[htb]
  \begin{picture}(10,6)(0,0)
\includegraphics{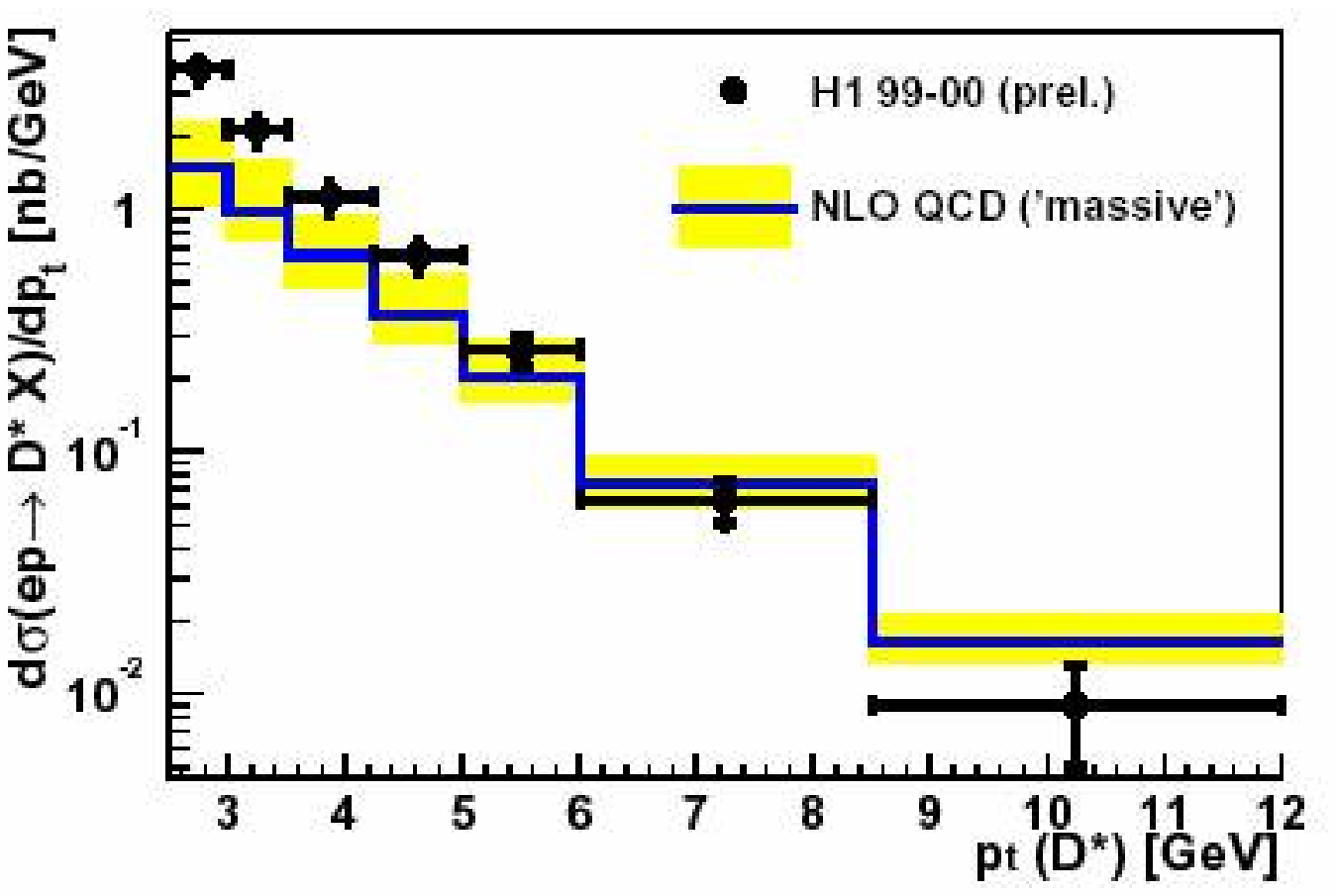}
\includegraphics{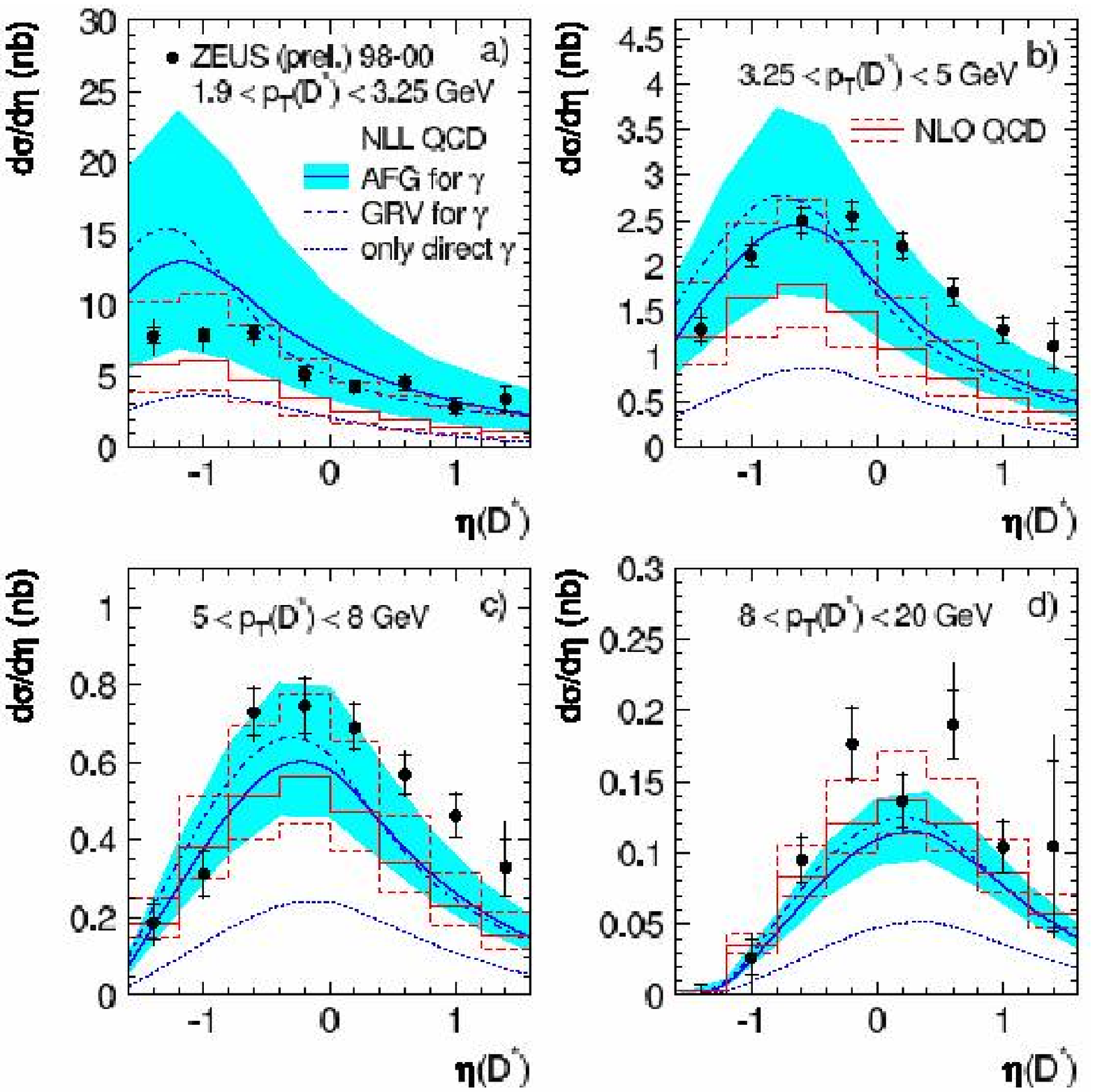}
\end{picture}
\caption{\label{fig:gpdst}
Charm meson cross section, measured by H1 and ZEUS at HERA,
compared to massive (labeled ``NLO'') 
and massless (labeled ``NLL'') NLO QCD calculations.}
\end{figure}

Several new results were reported from HERA.
H1 charm photoproduction cross sections 
have been updated using the full pre-upgrade luminosity~\cite{flucke}.
They rely on events with a tagged final state positron  
and have therefore smaller statistics than the ZEUS measurements.
They are found to be above the massive fixed order QCD predictions, again
particularly at low transverse momenta, see Fig.~\ref{fig:gpdst}.  
\unitlength1cm
The ZEUS data~\cite{hall-wilton}, also shown, 
have meanwhile a ``terrible precision'', 
as it may be seen from the theoretical point of view.
They are in principle sensitive to the charm density in the photon,
in the massless picture,  
however the calculations using different parton density parameterizations 
differ most at low $p_T$ 
where the theoretical uncertainties are much larger than these differences.

In DIS, a number of new results from their ongoing heavy quark production 
analyses were reported by CHORUS~\cite{dilellis}. 
For example, their data on charm meson production
in charged current neutrino nucleon scattering improve the experimental
constraints on the strange content of the nucleon. --   
H1 measured first differential jet cross sections~\cite{schmidt},
for dijet events produced in association with $D^*$ mesons.
Comparing with Monte Carlo predictions using different schemes for the 
evolution of the parton cascade, agreement is observed within expectation.  

\begin{figure}[htb]
  \begin{picture}(10,6)(0,0)
\includegraphics{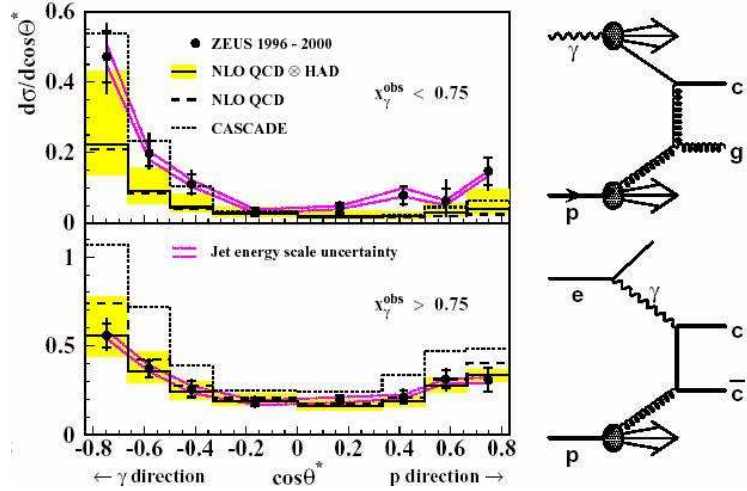}
\end{picture}
\caption{\label{fig:gpjets}
Angular distributions of charm tagged dijet events at HERA;
diagrams for charm excitation and boson gluon fusion.}
\end{figure}

The power of jet measurements is impressively demonstrated by 
the ZEUS analysis of charm dijet angular distributions in photoproduction,
which has recently been finalized~\cite{gladilin}.
The cross sections (Fig.~\ref{fig:gpjets}) are measured as a function of 
$\cos \theta^* = \tanh (\eta_{jet\, 1}-\eta_{jet\, 2})$
which can be reconstructed from the two jet rapidities and corresponds
to the polar angle of the outgoing parton in the partonic rest frame 
in a $2\ra 2$ process.
This is done separately for events with reconstructed $x_{\gamma}$
above or below 0.75, i.e. in regions where either direct or resolved 
photon contributions dominate.  
\unitlength1cm
First, the dependence on the absolute value of $\cos \theta^*$ 
reflects the shape of the hard QCD matrix element;
for low $x_{\gamma}$ the steep behaviour indicates the predominance of
gluon exchange in resolved processes, as also observed in inclusive jet data. 
Associating the $D^*$ meson with one of the jet 
tags a final state charm quark. 
The data show that for low $x_{\gamma}$ the latter is found predominantly 
in the photon direction (negative $\cos \theta^*$). 
This would not be expected in gluon gluon fusion 
(and is not seen in the direct, high $x_{\gamma}$ case dominated by
photon gluon fusion), 
but matches well the predictions based on diagrams with a 
$c$ quark originating from the hadronic photon structure in the 
initial state.
Such processes are included in the fixed order massive QCD calculation, 
but only as HO correction to direct photoproduction.
In the massless picture, this is a leading order process, and large NLO 
corrections are indeed found in the single inclusive case. 
One should also note the rather large uncertainty of the prediction 
in the massive scheme in this region.
This example shows how constraints in HERA kinematics and the quark hadron 
correlation in heavy meson production can be used to identify corners 
of phase space where improvements in the calculations are needed.

$b$ production at HERA is omitted in this write-up. 
Interesting discussions on the different cross section definitions of H1 and 
ZEUS and on their 
respective shortcomings took place~\cite{corradi,chiochia,gerlich}, 
but the overall picture changed with new H1 photoproduction 
results appearing at the
2003 summer conferences~\cite{H1b03},
with better precision and presented in a way allowing direct comparison to 
ZEUS. They show consistency between old and new data, and between experiments 
in the observable range; altogether they tend to be above expectations 
based on NLO QCD in the massive scheme. 

In hadroproduction, HERA B has appeared on the scene with a $b$ 
cross section measurement~\cite{braeuer}, which is in agreement with 
QCD predictions.
30 times more statistics has been collected, so one awaits 
the measurement of differential cross sections.

\begin{figure}[htb]
  \begin{picture}(10,5)(0,0)
\includegraphics{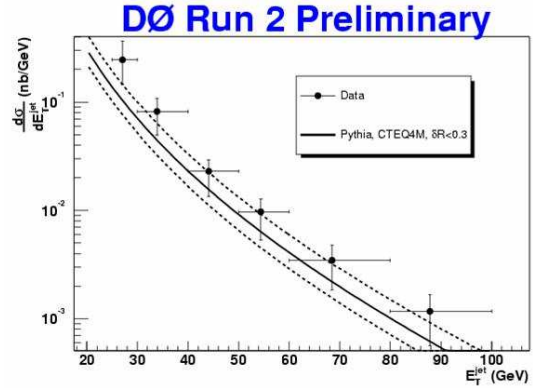}
\end{picture}
\caption{\label{fig:d0jets}
$b$ tagged jet cross section at the Tevatron.}
\end{figure}

Tagged $b$ jet cross sections are in principle less affected by 
theoretical uncertainties arising from collinear radiation than 
B meson or $b$ quark cross sections.  
The latest D0 data from run 2~\cite{khanov}
are consistent with run 1 results and are compared with 
MC predictions in Fig.~\ref{fig:d0jets}.
\unitlength1cm
  
The increased energy of the upgraded Tevatron implies an extra bonus 
of 30\% in the expected top production rate -- and the possibility 
to test its energy dependence. The first top cross section~\cite{gotra}
from run 2 was presented and found to be consistent with such expectations, 
within large errors still, but marking a first step towards QCD tests 
with differential top distributions.
 
\section{Conclusion}

As for heavy-quarkonium production, the NRQCD factorization approach
\cite{bbl} represents a rigorous theoretical framework that is renormalizable,
devoid of infrared singularities, and predictive.
It generalizes the CSM curing its defects.
However, NRQCD predictions for charmonium suffer from sizeable normalization
uncertainties at LO, and it is an important task for the future to evaluate
NLO corrections, both in $\alpha_s$ and $v^2$.
At LO, there is room for nonperturbative ingredients, such as $k_T$
factorization, hard-comover enhancement, and shape functions, which tend to
reduce the relative importance of the CO processes and ameliorate the
well-known shortages of NRQCD to describe the large-$z$ behaviour of inelastic
photoproduction at HERA and the charmonium polarization in hadroproduction at
large $p_T$ at the Tevatron.
The necessity for such nonperturbative effects is expected to be reduced as
NLO corrections are included.

As for open heavy-flavour production, the notorious excess of measurements of
inclusive $D$- and $B$-meson production in various experiments over the
respective FFNS analyses implemented with phenomenological FFs not
conceptually necessitated by factorization is expected to be reduced to an
acceptable level by adopting the GMVFNS, in which the massive quark also
appears in the initial state and would-be collinear singularities are
subtracted to match the $\overline{\rm MS}$ calculation for $p_T\gg m_Q$.
Then, the factorization theorem is at work, guaranteeing the universality and
DGLAP evolution of the (nonperturbative) FFs, and a global data analysis is
feasible.

From the experimental side, one notes that many new results continue to 
increase the variety of channels with which the QCD framework
of heavy flavour production is confronted. The Tevatron definitely 
presented itself as an upcoming charm, beauty and top factory. 
The gross picture for open heavy flavours is that theory undershoots,
in particular at low $p_T$, and that error bands 
in the predictions should be taken seriously, as they are in general
being exhausted once data appear. 
HERA analyses continue to explore new domains of precision and phase space;
they hold the potential to provide further guidance
to refine strategies in perturbative QCD,
in particular with the enhanced potential of the upgraded collider and 
detector coming to production mode this year~\cite{list}.

\section*{Acknowledgements} We thank all participants
for their contributions to a very lively and interesting session,
and the organizers for the invitation to a location
so rich of exceptional charm and beauty.


\begin{thebibliography}{99}

\bibitem{naroska} B. Naroska, these proceedings. 

\bibitem{tun} W.-K. Tung, S. Kretzer, and C. Schmidt,
J.\ Phys.\ G {\bf28}, 983 (2002);
W.-K. Tung, in these proceedings.

\bibitem{oln} F. Olness, in these proceedings.

\bibitem{bbl} G.T. Bodwin, E. Braaten, and G.P. Lepage,
Phys.\ Rev.\ D {\bf51}, 1125 (1995);
{\bf55}, 5853(E) (1997).

\bibitem{ber} V.G. Kartvelishvili, A.K. Likhoded, and S.R. Slabospitski\u i,
Yad.\ Fiz.\ {\bf28}, 1315 (1978) [Sov.\ J. Nucl.\ Phys.\ {\bf28}, 678 (1978)];
E.L. Berger and D. Jones,
Phys.\ Rev.\ D {\bf23}, 1521 (1981);
R. Baier and R. R\"uckl,
Phys.\ Lett.\ {\bf102B}, 364 (1981);
Z. Phys.\ C {\bf19}, 251 (1983).

\bibitem{bra} E. Braaten and S. Fleming,
Phys.\ Rev.\ Lett.\ {\bf74}, 3327 (1995);
P. Cho and A.K. Leibovich,
Phys.\ Rev.\ D {\bf53}, 150 (1996); {\bf53}, 6203 (1996).

\bibitem{abe} CDF Collaboration, F. Abe {\it et al.},
Phys.\ Rev.\ Lett.\ {\bf79}, 572 (1997); {\bf79}, 578 (1997);
D0 Collaboration, B. Abbott {\it et al.},
{\it ibid.}\ {\bf82}, 35 (1999).

\bibitem{bar} R. Barbieri, R. Gatto, and E. Remiddi,
Phys.\ Lett.\ {\bf61B}, 465 (1976).

\bibitem{zun} M. Kr\"amer, J. Zunft, J. Steegborn, and P.M. Zerwas,
Phys.\ Lett.\ B {\bf348}, 657 (1995);
M. Kr\"amer, Nucl.\ Phys.\ {\bf B459}, 3 (1996).

\bibitem{gammap} H1 Collaboration, C. Adloff {\it et al.},
Eur.\ Phys.\ J. C {\bf25}, 25 (2002);
ZEUS Collaboration, S. Chekanov {\it et al.},
{\it ibid.}\ {\bf27}, 173 (2003).

\bibitem{gg} M. Klasen, B.A. Kniehl, L.N. Mihaila, and M. Steinhauser,
Phys.\ Rev.\ Lett.\ {\bf89}, 032001 (2002).

\bibitem{pol} M. Klasen, B.A. Kniehl, L.N. Mihaila, and M. Steinhauser,
Phys.\ Rev.\ D {\bf68}, 034017 (2003);
M. Klasen and B.A. Kniehl, in these proceedings.

\bibitem{delphi} DELPHI Collaboration, J. Abdallah {\it et al.},
Phys.\ Lett.\ B {\bf565}, 76 (2003);
M. Chapkine, in these proceedings.

\bibitem{adl} PHENIX Collaboration, S.S. Adler {\it et al.},
hep-ph/0307019.

\bibitem{h1} H1 Collaboration, C. Adloff {\it et al.},
Eur.\ Phys.\ J. C {\bf10}, 373 (1999); {\bf25}, 41 (2002).

\bibitem{ep} B.A. Kniehl and L. Zwirner,
Nucl.\ Phys.\ {\bf B621}, 337 (2002).

\bibitem{cac} M. Cacciari and M. Kr\"amer,
Phys.\ Rev.\ Lett.\ {\bf76}, 4128 (1996);
P. Ko, J. Lee, and H.S. Song,
Phys.\ Rev.\ D {\bf54}, 4312 (1996); {\bf60}, 119902(E) (1999);
B.A. Kniehl and G. Kramer,
Eur.\ Phys.\ J. C {\bf6}, 493 (1999).

\bibitem{aff} CDF Collaboration, T. Affolder {\it et al.},
Phys.\ Rev.\ Lett.\ {\bf85}, 2886 (2000).

\bibitem{ben} M. Beneke and M. Kr\"amer,
Phys.\ Rev.\ D {\bf55}, 5269 (1997);
A.K. Leibovich,
{\it ibid.}\ {\bf56}, 4412 (1997).

\bibitem{bkl} E. Braaten, B.A. Kniehl, and J. Lee,
Phys.\ Rev.\ D {\bf62}, 094005 (2000);
B.A. Kniehl and J. Lee,
{\it ibid.}\ {\bf62}, 114027 (2000);
B.A. Kniehl, G. Kramer, and C.P. Palisoc,
Report No.\ DESY 03-096, hep-ph/0307386, Phys.\ Rev.\ D (in press).

\bibitem{sri} K. Sridhar, A.D. Martin, and W.J. Stirling,
Phys.\ Lett.\ B {\bf438}, 211 (1998);
Ph.\ H\"agler, R. Kirschner, A. Sch\"afer, L. Szymanowski, and O.V. Teryaev,
Phys.\ Rev.\ Lett.\ {\bf86}, 1446 (2001);
Phys.\ Rev.\ D {\bf63}, 077501 (2001);
F. Yuan and K.-T. Chao,
{\it ibid.}\ {\bf63}, 034006 (2001);
Phys.\ Rev.\ Lett.\ {\bf87}, 022002 (2001);
V.A. Saleev,
Phys.\ Rev.\ D {\bf65}, 054041 (2002);
V.A. Saleev and D.V. Vasin,
Phys.\ Lett.\ B {\bf548}, 161 (2002);
A.V. Lipatov and N.P. Zotov,
Eur.\ Phys.\ J. C {\bf27}, 87 (2003);
A.V. Kotikov, in these proceedings;
V.A. Saleev, in these proceedings;
N.P. Zotov, in these proceedings.

\bibitem{spb} S.P. Baranov,
Phys.\ Rev.\ D {\bf66}, 114003 (2002).

\bibitem{hoy} P. Hoyer and S. Peign\'e,
Phys.\ Rev.\ D {\bf59}, 034011 (1999);
N. Marchal, S. Peign\'e, and P. Hoyer,
{\it ibid.}\ {\bf62}, 114001 (2000);
S. Peign\'e, in these proceedings.

\bibitem{cem} H. Fritzsch,
Phys.\ Lett.\ {\bf67B}, 217 (1977);
F. Halzen,
{\it ibid.}\ {\bf69B}, 105 (1977);
M. Gl\"uck, J.F. Owens, and E. Reya,
Phys.\ Rev.\ D {\bf17}, 2324 (1978).

\bibitem{sch} G.A. Schuler and R. Vogt,
Phys.\ Lett.\ B {\bf387}, 181 (1996);
J.F. Amundson, O.J.P. Eboli, E.M. Gregores, and F. Halzen,
{\it ibid.}\ B {\bf390}, 323 (1997).

\bibitem{katkov} I. Katkov, in these proceedings. 

\bibitem{kruecker} D.Kr\"ucker, in these proceedings. 

\bibitem{rvogtadep}
R.~Vogt,
Nucl.\ Phys.\ A {\bf 700} (2002) 539
[arXiv:hep-ph/0107045].

\bibitem{pakhlov} P. Pakhlov, in these proceedings. 

\bibitem{meyer} A. Meyer, in these proceedings. 

\bibitem{kkp} B.A. Kniehl, G. Kramer, and B. P\"otter,
Nucl.\ Phys.\ {\bf B582}, 514 (2000); {\bf B597}, 337 (2001);
S. Kretzer,
Phys.\ Rev.\ D {\bf62}, 054001 (2000);
L. Bourhis, M. Fontannaz, J.P. Guillet, and M. Werlen,
Eur.\ Phys.\ J. C {\bf19}, 89 (2001).

\bibitem{bkkc} J. Binnewies, B.A. Kniehl, and G. Kramer,
Phys.\ Rev.\ D {\bf58}, 014014 (1998).

\bibitem{bkkb} J. Binnewies, B.A. Kniehl, and G. Kramer,
Phys.\ Rev.\ D {\bf58}, 034016 (1998);
G. Kramer, in these proceedings.

\bibitem{col} J.C. Collins,
Phys.\ Rev.\ D {\bf58}, 094002 (1998);
private communication.

\bibitem{pet} C. Peterson, D. Schlatter, I. Schmitt, and P.M. Zerwas,
Phys.\ Rev.\ D {\bf27}, 105 (1983).

\bibitem{pff} J. Binnewies, B.A. Kniehl, and G. Kramer,
Z. Phys.\ C {\bf76}, 677 (1997).

\bibitem{gre} M. Cacciari and M. Greco,
Phys.\ Rev.\ D {\bf55}, 7134 (1997).

\bibitem{cn} M. Cacciari and P. Nason,
Phys.\ Rev.\ Lett.\ {\bf89}, 122003 (2002).

\bibitem{acot} M.A.G. Aivazis, J.C. Collins, F.I. Olness, and W.-K. Tung,
Phys.\ Rev.\ D {\bf50}, 3102 (1994);
F.I. Olness, R.J. Scalise, and W.-K. Tung,
{\it ibid.}\ {\bf59}, 014506 (1999);
A. Chuvakin, J. Smith, and W.L. van Neerven,
{\it ibid.}\ {\bf61}, 096004 (2000).

\bibitem{ks1} G. Kramer and H. Spiesberger,
Eur.\ Phys.\ J. C {\bf22}, 289 (2001);
H. Spiesberger, in these proceedings.

\bibitem{ks2} G. Kramer and H. Spiesberger,
Eur.\ Phys.\ J. C {\bf28}, 495 (2003).

\bibitem{cdf} CDF Collaboration, F. Abe {\it et al.},
Phys.\ Rev.\ Lett.\ {\bf75}, 1451 (1995);
CDF Collaboration, D. Acosta {\it et al.},
Phys.\ Rev.\ D {\bf65}, 052005 (2002).

\bibitem{nas} P. Nason, S. Dawson, and R.K. Ellis,
Nucl.\ Phys.\ {\bf B327}, 49 (1989); {\bf B335}, 260(E) (1990);
W. Beenakker, H. Kuijf, W.L. van Neerven, and J. Smith,
Phys.\ Rev.\ D {\bf40}, 54 (1989).

\bibitem{kerzel} U. Kerzel, in these proceedings. 

\bibitem{landsman} H. Landsman, in these proceedings. 

\bibitem{harder} 
K.~Harder,
``B fragmentation and energy correlation in Z $\to$ B anti-B decays (LEP-1, SLD),''
{\it Prepared for 31st International Conference on High Energy Physics (ICHEP 2002), Amsterdam, The Netherlands, 24-31 Jul 2002}

\bibitem{gladilin} L. Gladilin, in these proceedings. 

\bibitem{giammanco} A. Giammanco, in these proceedings. 

\bibitem{achard} P. Achard, in these proceedings. 

\bibitem{flucke} G. Flucke, in these proceedings. 

\bibitem{hall-wilton} R. Hall-Wilton, in these proceedings. 

\bibitem{dilellis} G. De Lellis, in these proceedings. 

\bibitem{schmidt} S. Schmidt, in these proceedings. 

\bibitem{corradi} M. Corradi, in these proceedings. 

\bibitem{chiochia} V. Chiochia, in these proceedings. 

\bibitem{gerlich} C. Gerlich, in these proceedings. 

\bibitem{H1b03} O. Behnke, prepared for EPS conference, Aachen, July 2003. 

\bibitem{braeuer} M. Br\"auer, in these proceedings. 

\bibitem{khanov} A. Khanov, in these proceedings. 

\bibitem{gotra} Yu.\ Gotra, in these proceedings. 

\bibitem{list} B. List, in these proceedings. 

\end{thebibliography}
\end{document}